\documentclass[review]{elsarticle}

\usepackage{lineno,hyperref,amsmath,graphicx,amssymb,subfig,rotating}
\modulolinenumbers[5]

\journal{Epidemics}

\begin{document}

\begin{frontmatter}

 \title{A new logistic growth model applied to COVID-19 fatality data}
\author[uwc]{S.~Triambak}
\ead{striambak@uwc.ac.za}

\author[utkal]{D.\,P.~Mahapatra}
\ead{dpm.iopb@gmail.com}

\author[iit]{N.~Mallick}
\author[iit]{R.~Sahoo\fnref{fn1}}
\fntext[fn1]{Present address: CERN, Esplanade des Particules, Meyrin, Switzerland}
\address[uwc]{Department of Physics and Astronomy, University of the Western Cape, P/B X17, Bellville 7535, South Africa}
\address[utkal]{Department of Physics, Utkal University, Vani Vihar, Bhubaneshwar 751004, India}
\address[iit]{Department of Physics, Indian Institute of Technology Indore, Simrol, Indore 453552, India}

\begin{abstract}
\textit{Background:} Recent work showed that the temporal growth of the novel coronavirus disease (COVID-19) follows a sub-exponential power-law scaling whenever effective control interventions are in place. Taking this into consideration, we present a new phenomenological logistic model that is well-suited for such power-law epidemic growth.\\ 
\textit{Methods:} We empirically develop the logistic growth model using simple scaling arguments, known boundary conditions and a comparison with available data from four countries, Belgium, China, Denmark and Germany, where (arguably) effective containment measures were put in place during the first wave of the pandemic. A non-linear least-squares minimization algorithm is used to map the parameter space and make optimal predictions.\\ 
\textit{Results:} Unlike other logistic growth models, our presented model is shown to consistently make accurate predictions of peak heights, peak locations and cumulative saturation values for incomplete epidemic growth curves. We further show that the power-law growth model also works reasonably well when containment and lock down strategies are not as stringent as they were during the first wave of infections in 2020.  On the basis of this agreement, the model was used to forecast COVID-19 fatalities for the third wave in South Africa, which is currently in progress.
\\   
\textit{Conclusions:} We anticipate that our presented model will be useful for a similar forecasting of COVID-19 induced infections/deaths in other regions as well as other cases of infectious disease outbreaks, particularly when power-law scaling is observed.   
\end{abstract}

\end{frontmatter}


\section{Introduction}
\label{intro}

The COVID-19 pandemic has reinvigorated efforts at an unprecedented scale to better understand the dynamics and mechanism of infectious disease spread. Presently, there is  significant interest worldwide to model region-specific infection and mortality curves, while also working on effective intervention and containment strategies. It is hoped that such a collective endeavor would continue working towards preventing an uncontrolled proliferation of the disease, while simultaneously countering near irreparable socio-economic damage from multiple waves of infections. This has resulted in a deluge of scientific literature related to the pandemic, that have proved to be a challenge to keep up with~\cite{tsunami}. A large subset of  research papers investigated the spatio-temporal evolution of the disease~\cite{chinazzi,Gatto,Gross}, mostly using variants of the compartmental SIR (Susceptible-Infected-Removed) epidemiological model~\cite{maier,axel,barman,wu,bust,li2020,roques} to analyze the number of infections (or deaths) in specific regions. Other methods involved the use of phenomenological models~\cite{maimuna,Roosa}, time-varying and non-linear Markov processes~\cite{ziqi,markov}, superpositions of epidemic waves~\cite{koltsova}, hybrid nonparametric models~\cite{wang} and other data-driven approaches~\cite{salas,schneble,altmejd}, including those based on artificial intelligence~\cite{ai}, etc. Along these lines, we recently performed a random walk Monte Carlo study to make temporal growth exponent predictions for COVID-19-like disease spread~\cite{st}, particularly for a spatially constrained, yet stochastically interacting population. In that work, similar to other simulational approaches~\cite{Mollison,Filipe1}, the spread of the disease was modeled on the basis of proximity-based interactions. We identified certain similarities between our simulation results and those obtained from other compartmentalized differential-equation-based extended SIR models~\cite{maier,axel}, and further showed that spatial mobility plays a key role in determining the eventual growth in the total number of infections/deaths as a function of time. While this conclusion should not be surprising~\cite{fisher}, it was corroborated by a recent data-driven analysis of the `mobility-network' in Germany, using cellular phone data~\cite{schlosser}. These investigations established a connection between the three approaches (data-driven, simulation, and compartmental model) used to better understand infectious disease spread.

In terms of phenomenological modelling, 
some of the most extensively used tools for making COVID-19 related predictions are the family of logistic growth models (LGMs)~\cite{Roosa,masjedi,wu20,singer,batista,vattay,morais,shen,yang,jia,dattoli,sonnino,molina}. This is not unexpected. LGMs have been successfully used in the past for predicting growth curves in epidemics such as Ebola, SARS, H1N1, dengue, etc.~\cite{pell,chowell,wang12}. 
Along these lines, our previous work~\cite{st} also showed that the commonly used generalized LGM~\cite{richards} works for exponential growth, it may fail to satisfactorily fit epidemic data when there is sub-exponential power-law growth. As a continuation of our engagement with this problem, we present in this work a logistic growth model that  describes such data more accurately.    
\section{Logistic growth models}
\label{lgms}
%
 In the context of the COVID-19 pandemic, the simplest LGM used by some research groups~\cite{singer,batista,vattay,morais,shen} is described by the well-known Verhulst differential equation 
\begin{equation}
\label{eq:verhulst}
\frac{dN}{dt} = \lambda N \left(1-\frac{N}{K}\right),
\end{equation}
where $\lambda$ is the intrinsic growth constant and $K$ (also called the carrying capacity) is the asymptotic (saturation) limit for $N(t)$, as $t \to \infty$. The general solution of Eq.~\eqref{eq:verhulst} is of the form
\begin{equation}                                                                                                                                                                      
\label{eq:classical}
N(t) = \frac{K}{1 + B \exp{(-\lambda t)}},                                                                                                                                                                      \end{equation} 
where 
the point of inflection is at $ t = \ln B/\lambda$. The above is a special case of the Richards LGM~\cite{richards} 
\begin{equation}
N(t) = K \left[1 + B \exp(-\lambda t)\right]^{1/(1-m)}, 
\end{equation}
which is a solution of the Richards differential equation
\begin{equation}
\label{eq:richards}
\frac{dN}{dt} = \frac{\lambda N}{1-m}\left[\left(\frac{K}{N}\right)^{1-m} - 1\right]. 
\end{equation}
Here, the parameter $m$ decides both the shape of the growth curve as well as its inflection point. For example, as $m \to 1$, Eq.~\eqref{eq:richards} becomes the Gompertz growth curve~\cite{Gompertz,Laird}. The special case of $m = 2$ describes classical logistic growth, shown in Eqs.~\eqref{eq:verhulst} and \eqref{eq:classical}. To be consistent with other recent literature and given the fact that we are only interested in the family of curves with $m > 1$, we rewrite Eq.~\eqref{eq:richards} as 
\begin{equation}
\label{eq:richards2}
\frac{dN}{dt} = \lambda' N\left[1 - \left(\frac{N}{K}\right)^{q}\right], 
\end{equation}
where $q = \lvert1-m\rvert$ and $\lambda' = \lambda/q$.

%

Recently, it has been shown~\cite{chowell2}  
that in order to allow for sub-exponential growth, one can further generalize the Richards equation by replacing $N$ with $N^p$ in Eq.~\eqref{eq:richards2}, where $p \le 1$ is a `deceleration' parameter~\cite{viboud}.
Such sub-exponential growth was observed with initial COVID-19 data from China, where an analysis~\cite{maier} of the number of reported cases from several provinces in the country showed a $t^\alpha$ type power-law growth in $N$. This was attributed to effective containment and mitigation measures, as well as behavioral changes of the population~\cite{maier}. Such control interventions prevent a homogeneous mixing~\cite{Fofana} of the population, which if unchecked would lead to exponential growth, provided there is no depletion of the susceptible population~\cite{Bailey_book}. Our simulations~\cite{st} further showed that the minimum growth exponent obtained under the most stringent mobility restrictions is quadratic growth $(\alpha = 2)$. More realistically one would expect growth exponents that are slightly higher than 2, under the most effective containment scenarios~\cite{maier,st}. It is reasonable to expect that during the first wave of the pandemic (in 2020) most countries followed similar contaiment strategies at various levels to counter the spread of COVID-19 within their populace. Therefore their cumulative infection (and fatality) curves are expected to have power-law growth exponents in the range of 2 to 3~\cite{st}. Below we develop a new LGM that can adequately describe such data, and make reasonably accurate and consistent predictions. 
\section{Methods}
\subsection{Development of the power-law LGM}
To develop the model, we start similarly as in Eq.~\eqref{eq:verhulst}, with the ansatz that the daily infection rate is proportional to $N$, the number of individuals who are already infected by the disease. Furthermore, it is apparent that for bounded (logistic) growth, one requires the daily rate to also be proportional to a term similar to the ones described in the parentheses of Eqs.~\eqref{eq:verhulst} and \eqref{eq:richards2}. Therefore, for power-law growth, with $N \propto t^\alpha$, we write a general form of daily infection rate, analogous to Eq.~\eqref{eq:richards2} as 
\begin{equation}
\label{eq:ours1}
\frac{dN}{dt} = \lambda t^\alpha \left[1 - \left(\frac{t}{\beta}\right)^\gamma\right]^{\delta}. 
\end{equation}
It is important to note that here $dN/dt$ has an explicit dependence on time, unlike Eqs.~\eqref{eq:verhulst} and \eqref{eq:richards2}. The parameter $\beta$ is in units of time, so that in the asymptotic limit as $t \to \beta$ (well past the peak of the epidemic curve, for large values of $\beta$), $dN/dt \to 0$. The $\alpha$, $\gamma$ and $\delta$ parameters are dimensionless, while $\lambda$ has units of $1/t$.

In the next step, we empirically tested and developed this model, by fitting the above function to available data from four countries, Belgium, China, Denmark and Germany, during the first wave of infections in 2020. These countries were chosen because the data show a reasonably successful containment of the spread of COVID-19 within their population~\cite{WHO2020}, during the first wave. Similar to our previous work~\cite{st}, we performed a time-series analyses for the number of reported daily deaths\footnote{All data described in this work are 5-day rolling averaged.}, instead of infections. This was due to several reasons. Firstly, the death toll is far more important to quantify than the infection rate in a given population, although they are related. Secondly, when performing a global comparison of data from different countries, we assume that COVID-19-related deaths are more accurately and uniformly recorded \textit{in general}. And finally, given the strong correlation between the number of infected cases and number of deaths, the time-series trends in both death and infection rates are expected to be similar to one another.\footnote{We caution that one must be careful in making this assumption, which may fail when live-saving treatment options are put in place (or inaccessible) midway, thereby affecting daily mortality rates. Such real-time interventions affect all phenomenological models.}      

\subsection{Analysis}

The fits were performed using a non-linear least squares (NLS) algorithm that minimized the sum of squared residuals (SSR), defined by 
\begin{equation}
{\rm SSR} = \sum_{i = 1}^{t_\textrm{days}} \left[D_i-y(t_i)\right]^2, 
\end{equation}
with respect to the daily reported deaths $D_i$, 
where 
\begin{equation}
y(t_i) = \left(\frac{dN}{dt}\right)_{t_i},
\end{equation}
and $N$ is the cumulative number of deaths at time $t$. 
However, the NLS fitting procedure showed that the five parameter fit in Eq.~\eqref{eq:ours1} was not optimal for such analysis. The five parameters were found to be highly correlated, with correlation coefficients in the range of $0.83 \leq |\rho| \leq 1$. Successive fits to the same data, for different initial values for the parameters resulted in arbitrary and widely-varying values for the converged fit parameters, particularly $\lambda$, $\beta$ and $\delta$. Despite this, the fits yielded very similar values for the minimum SSR and nearly indistinguishable results. The above showed that the model in Eq.~\ref{eq:ours1} was not feasible for an out-of-sample forecasting with partial $dN/dt$ epidemic curves. Consequently, we modified Eq.~\eqref{eq:ours1} to 
\begin{equation}
\label{eq:ours2}
\frac{dN}{dt} = \lambda t^\alpha \left[1 - \left(\frac{t}{\beta}\right)^\gamma\right]^{\beta/\epsilon},  
\end{equation}
as a means to bypass the problem. For this part of the analysis, the $\beta$ parameter was kept fixed at a large value ($\beta = 500$~days). The above prescription reduced the problem to four parameters, while placing significant restrictions on the allowed parameter space. Similar fits, performed as before showed nearly no dependence on $\beta$ (as long as it is large and fixed), with the parameters consistently converging to very similar values.     

 \section{Results and discussion}
\subsection{Results for first wave data from 2020}
\begin{figure}[htb]
\centering
\subfloat{
\includegraphics[width=0.6\textwidth]{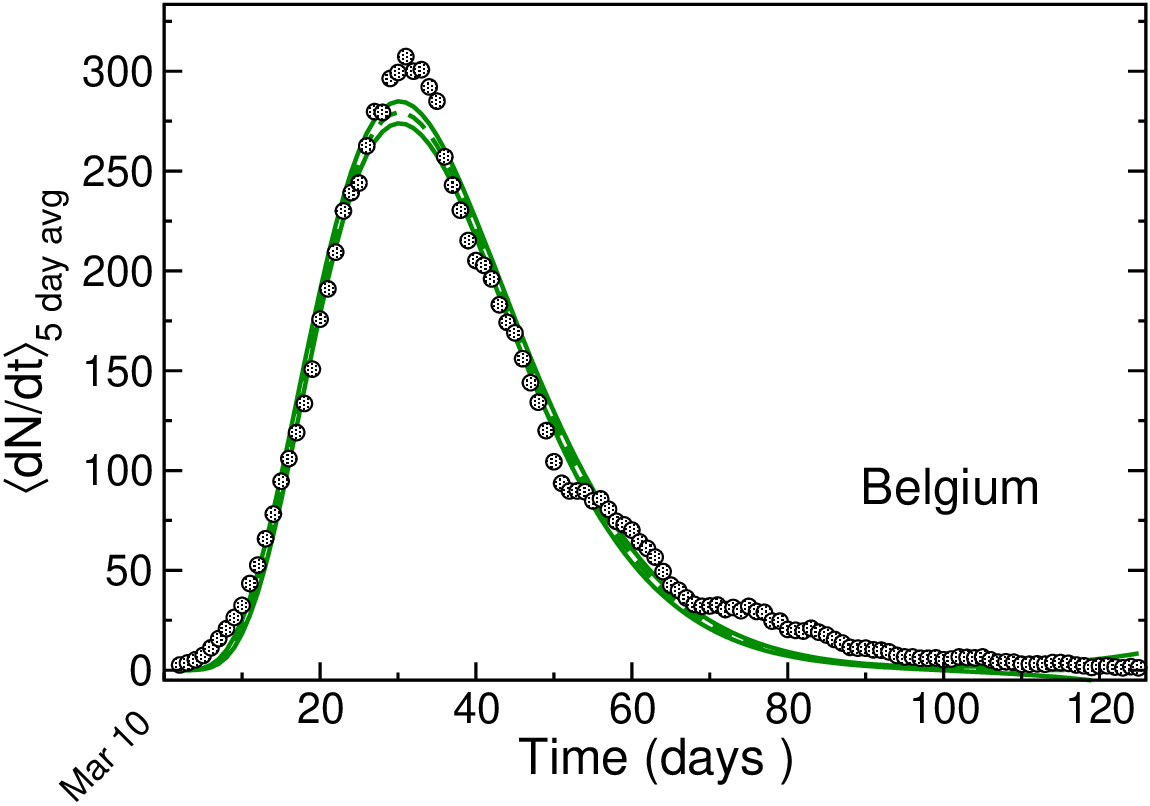}
}
\subfloat{
 \includegraphics[width=0.6\textwidth]{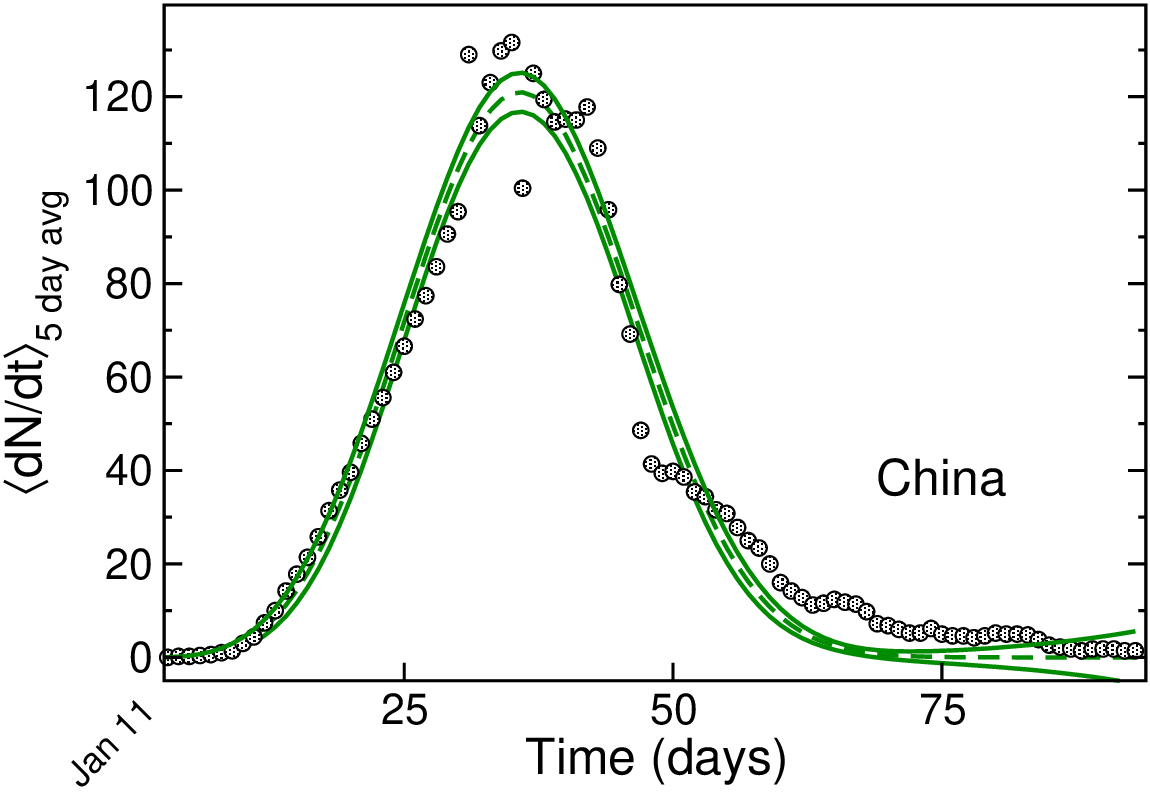}
 }
\vspace{0.5em}
\subfloat{
  \includegraphics[width=0.6\textwidth]{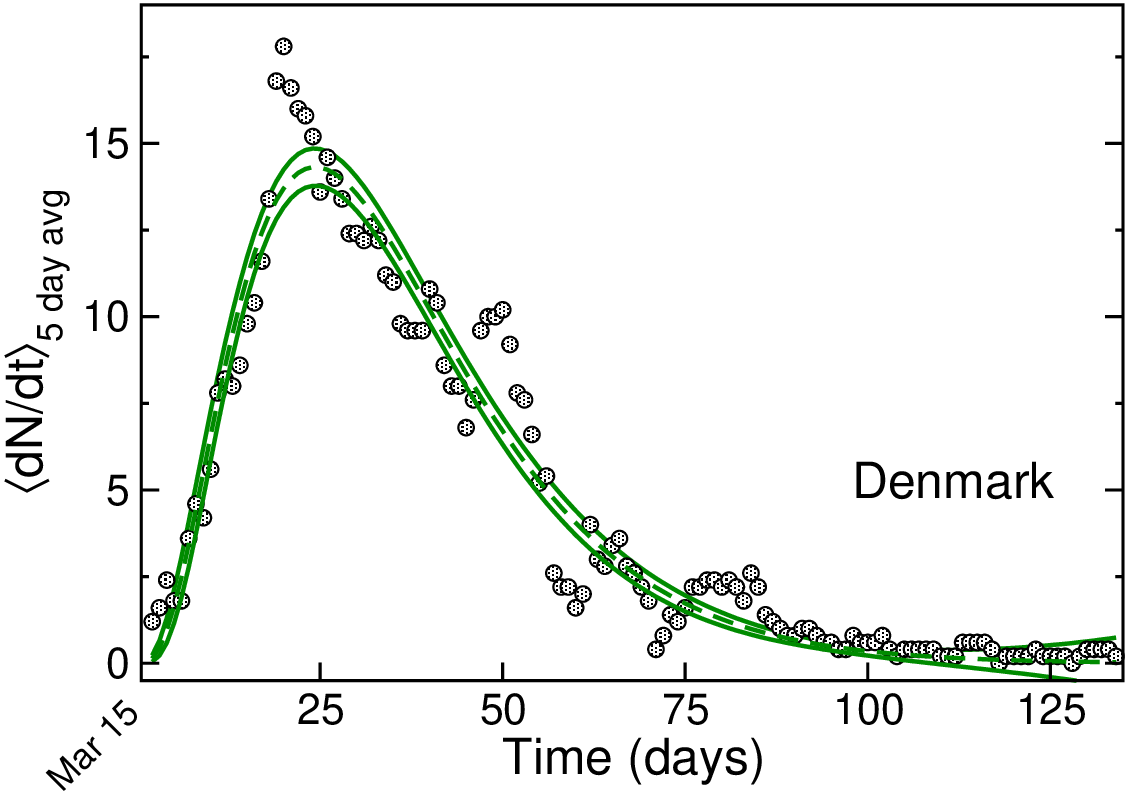}
}
\subfloat{
   \includegraphics[width=0.6\textwidth]{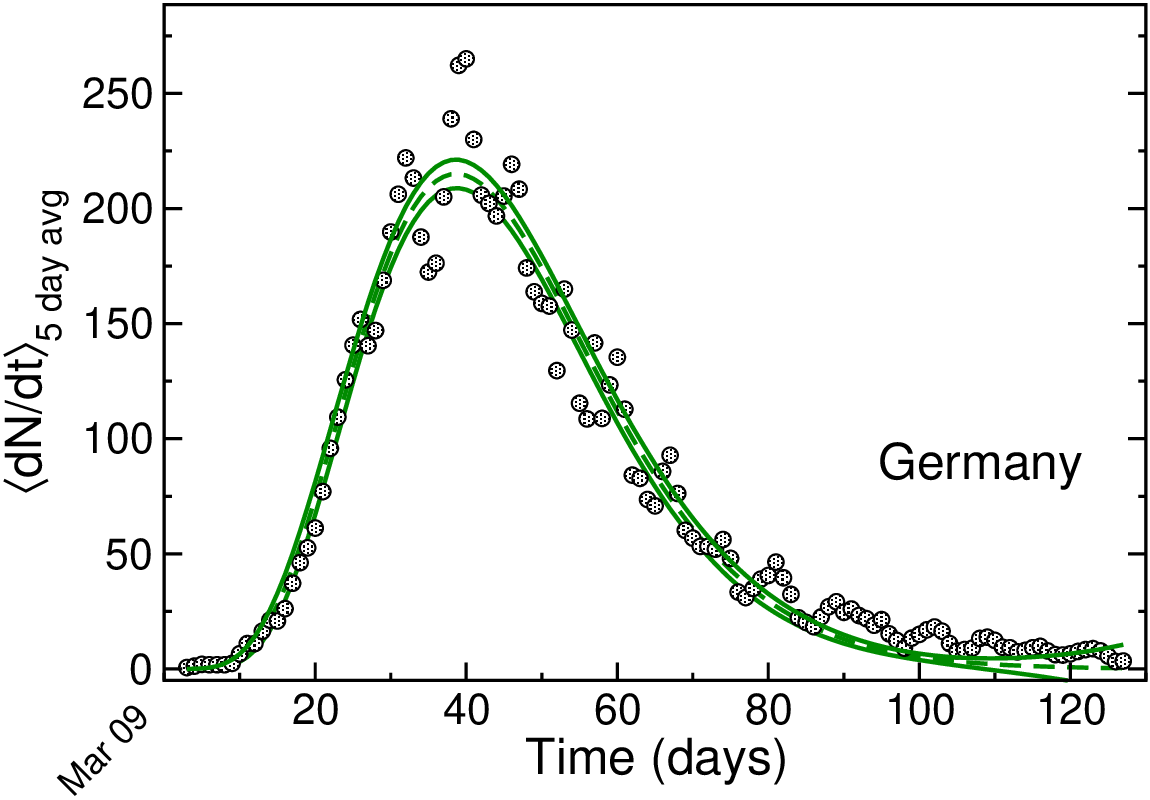}
   }
 \caption{\label{fig:full}Power-law growth model fits to reported deaths for Belgium, China, Denmark and Germany, shown with $\pm$95\% confidence interval (CI) bands (shown in green). The data points (filled circles) are generated from a 5 day moving average of number of daily deaths reported by the World Health Organization~\cite{WHO2020}, for the first wave of infections in 2020. In each case, the date of the first reported death (day 1) is indicated at $t = 0$ on the time axis.}
 \end{figure}

On fitting the data using our modified power-law logistic function,   
we obtain good agreement with the daily mortality curves from the four countries considered earlier. This is shown
in Fig.~\ref{fig:full}. As expected, the product of $\lambda$ and $t^\alpha$ mainly contribute to the rising part of the $dN/dt$ curve. The other two parameters $\gamma$ and $\epsilon$ contribute to truncating the rise. Together, these parameters decide the position of the central value of the peak as well as the nature of any tailing feature that follows it. The tailing in the data depends on country-specific mitigation and containment measures. It is found to be more prominent in the cases of Belgium, Denmark, and Germany. 
\begin{figure}[htb]
\centering
\subfloat{
\includegraphics[width=0.51\textwidth]{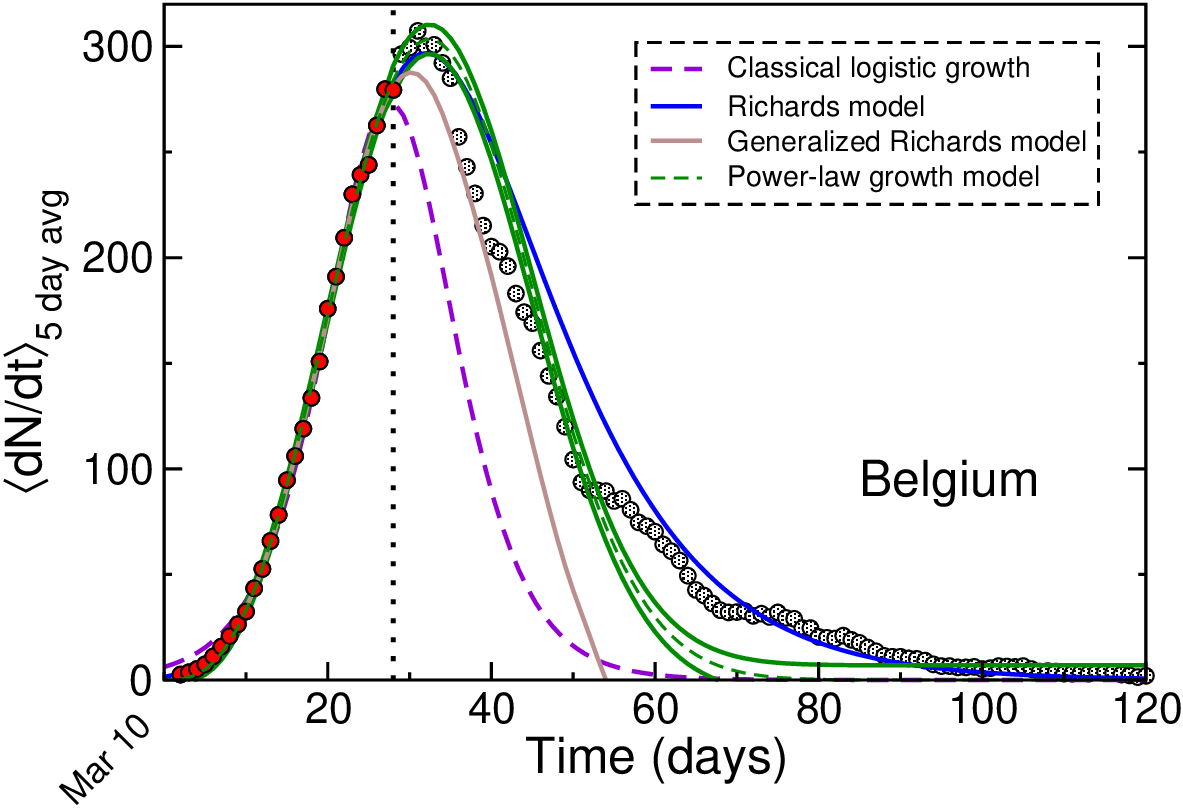}
}
\subfloat{
 \includegraphics[width=0.51\textwidth]{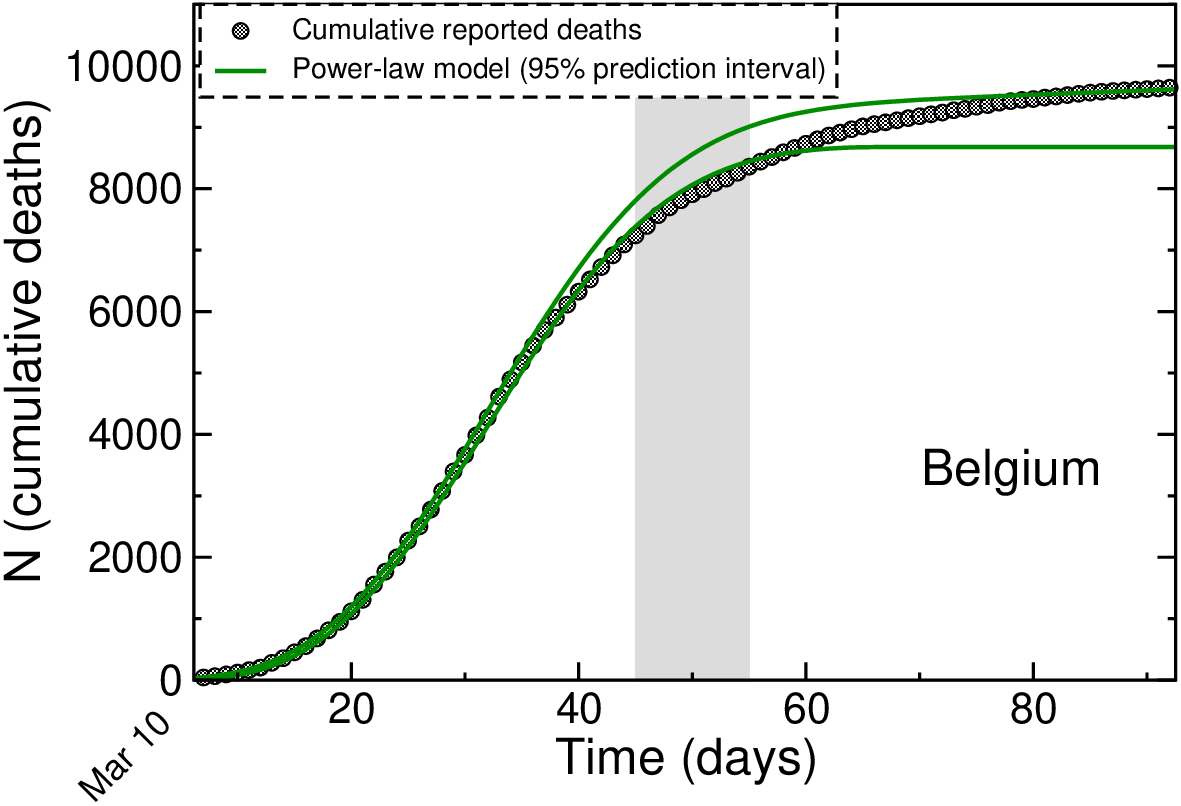}
 }
 \vspace{0.5em}
 \subfloat{
  \includegraphics[width=0.51\textwidth]{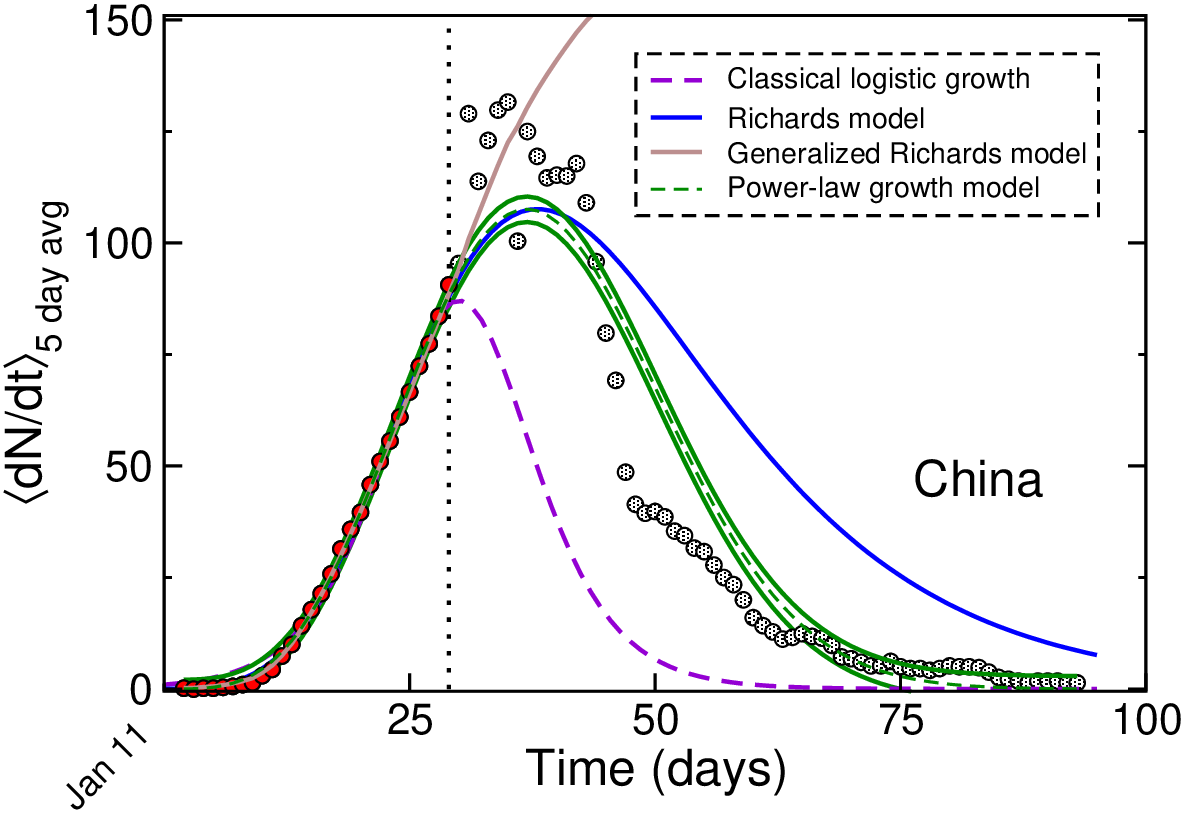}
  }
  \subfloat{
 \includegraphics[width=0.51\textwidth]{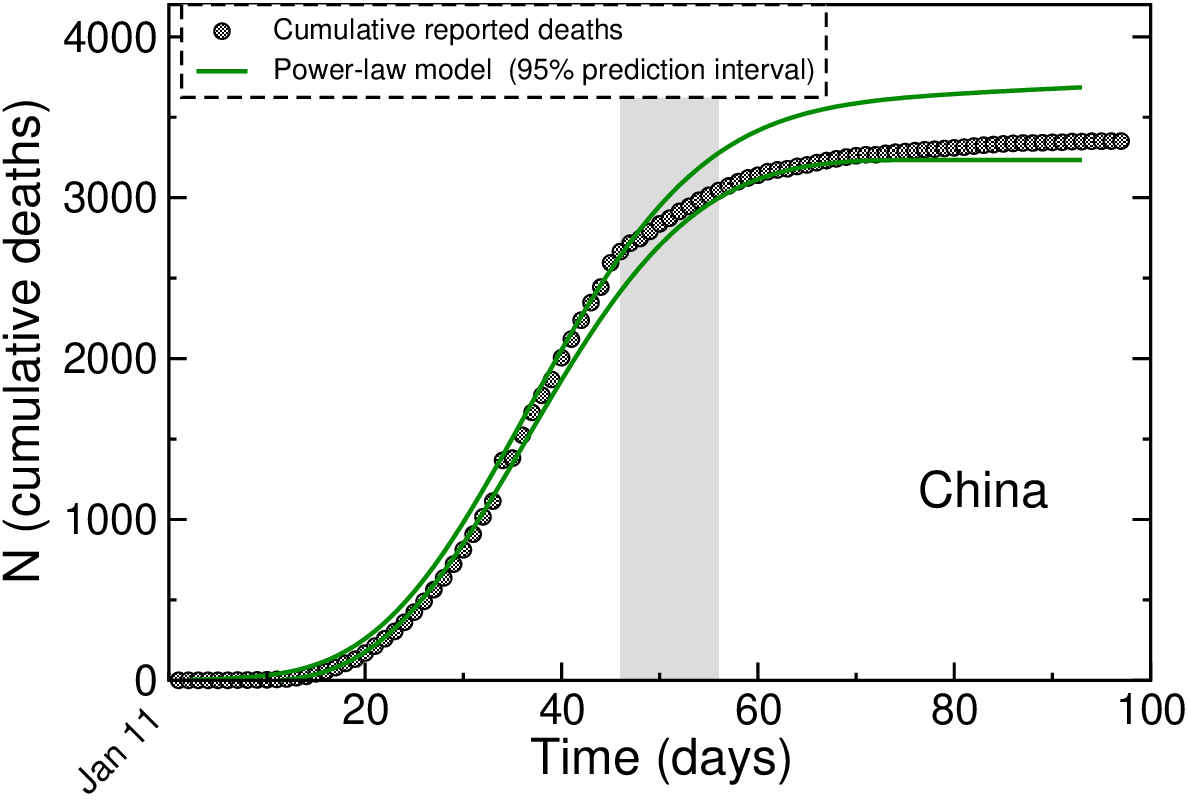}
 }
 \vspace{0.5em}
 \subfloat{
 \includegraphics[width=0.51\textwidth]{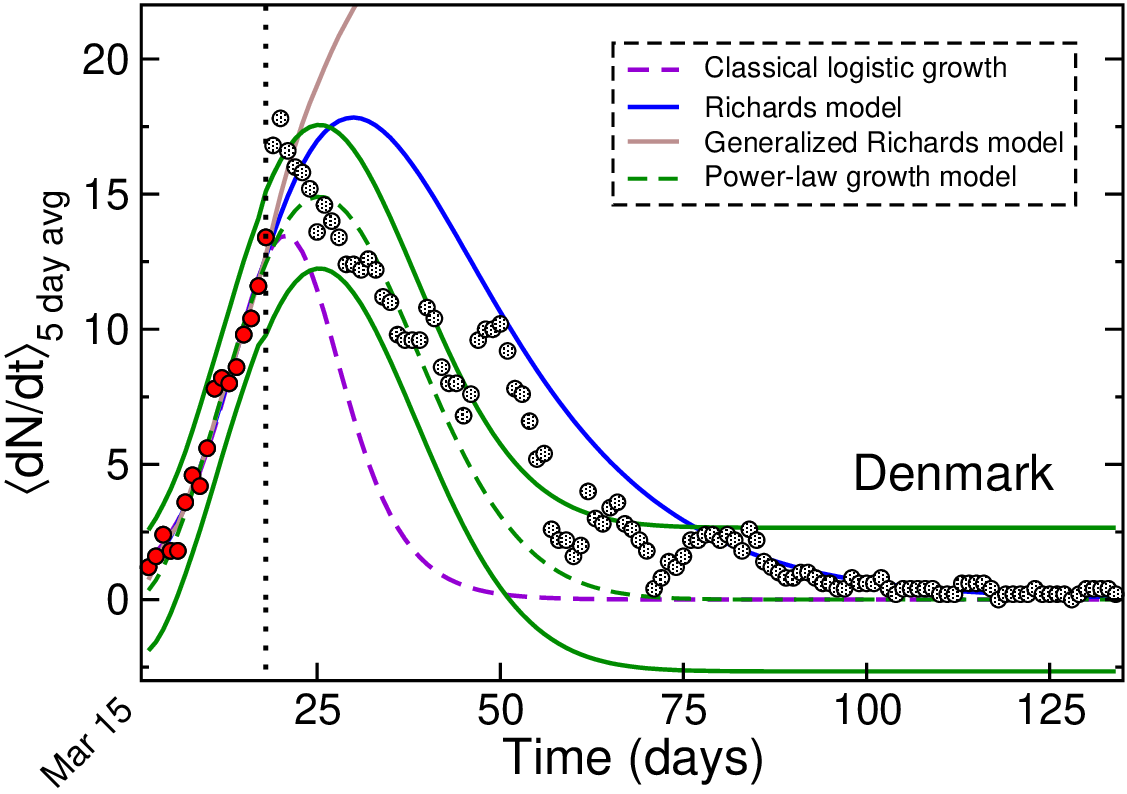}
 }
 \subfloat{
 \includegraphics[width=0.51\textwidth]{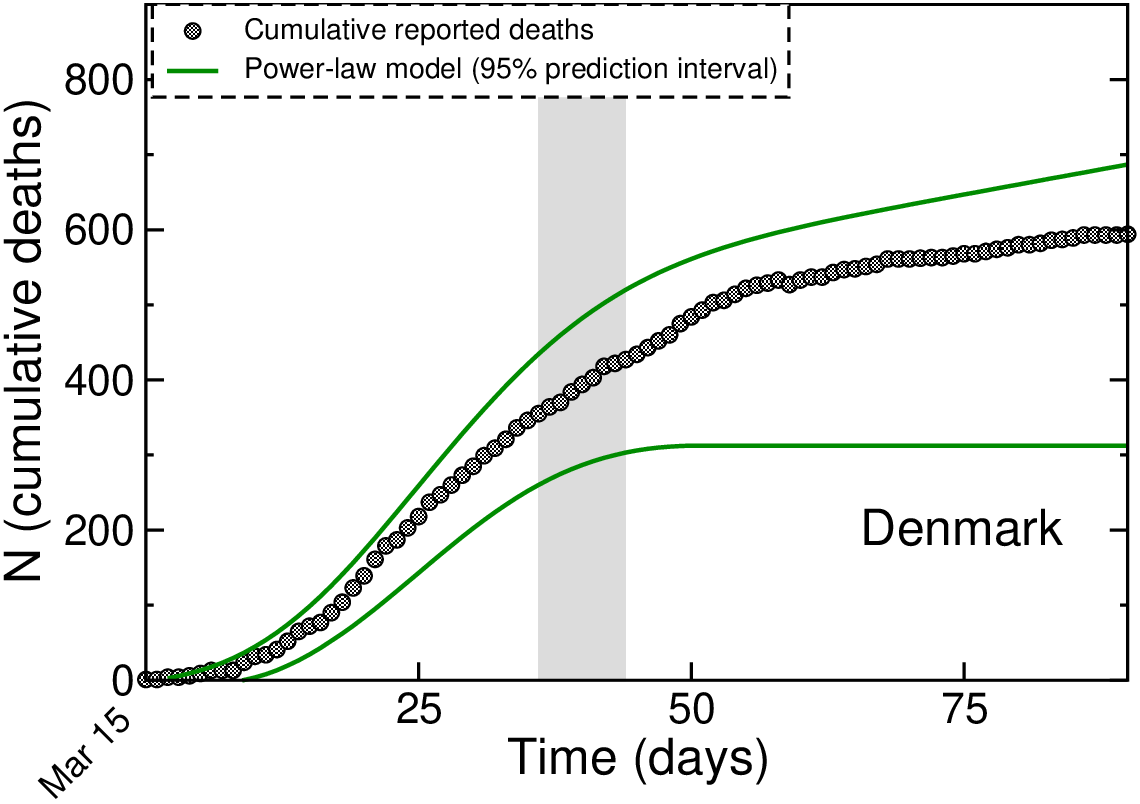}
 }
 \vspace{0.5em}
 \subfloat{
 \includegraphics[width=0.51\textwidth]{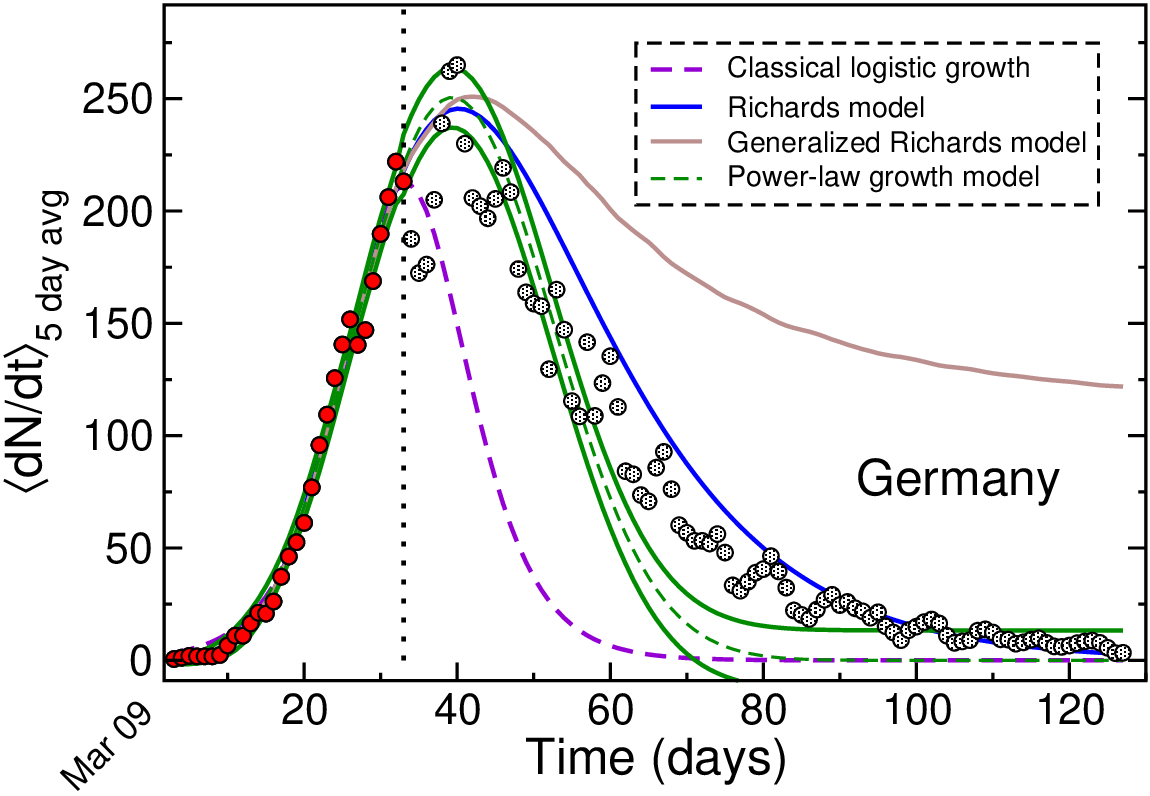}
 }
 \subfloat{
 \includegraphics[width=0.51\textwidth]{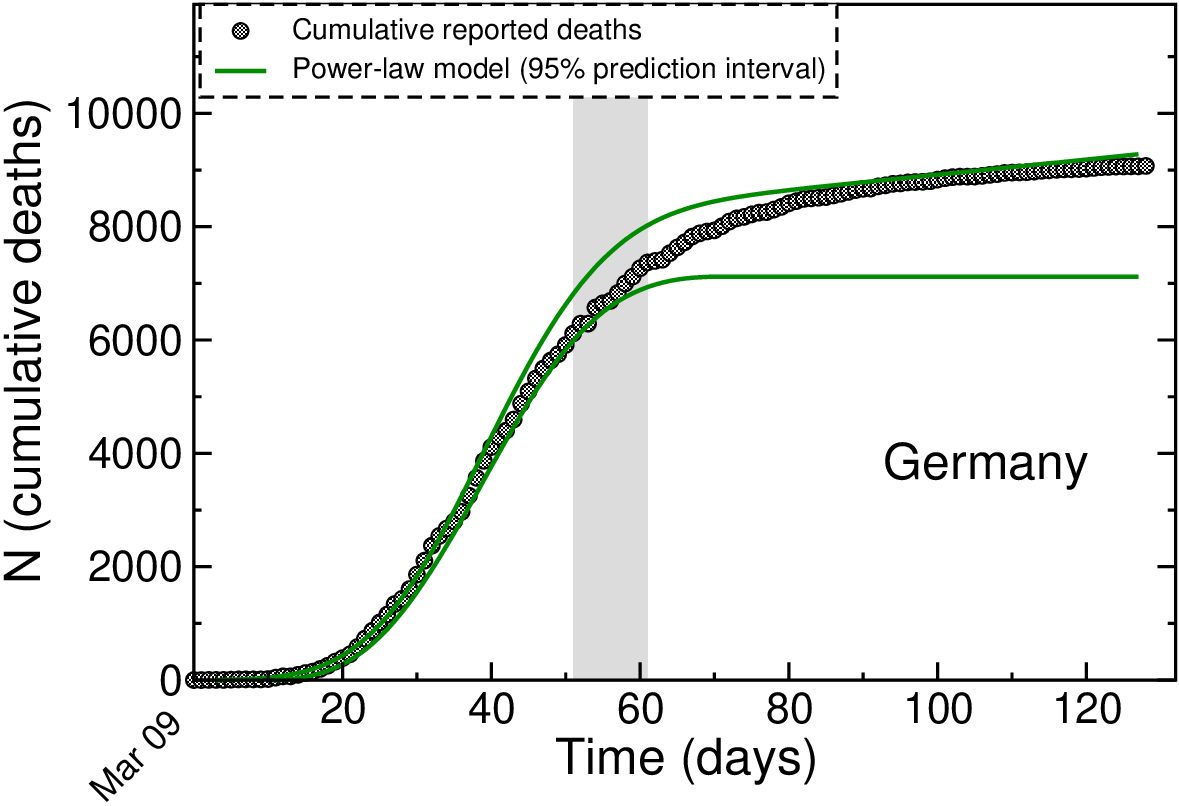}
 }
 \caption{\label{fig:rising}Left panel: Various LGM fits to the daily death data from Belgium, China, Denmark and Germany, shown together with 95\% prediction intervals for the power-law model. The data points are the same as shown in Fig.~\ref{fig:full} and correspond to the first waves in 2020. The red filled circles represent the in-sample calibration points used for the fits.
 Right panel: Cumulative data reconstructed from power-law growth model fits, shown together their 95\% prediction intervals. The grey band shows the approximate date range when the curves begin to flatten out. In each case, the date of the first reported death (day 1) is indicated at $t = 0$ on the time axis.}
  \end{figure}
\begin{figure}[htpb]
\centering
\subfloat{
\includegraphics[width=0.6\textwidth]{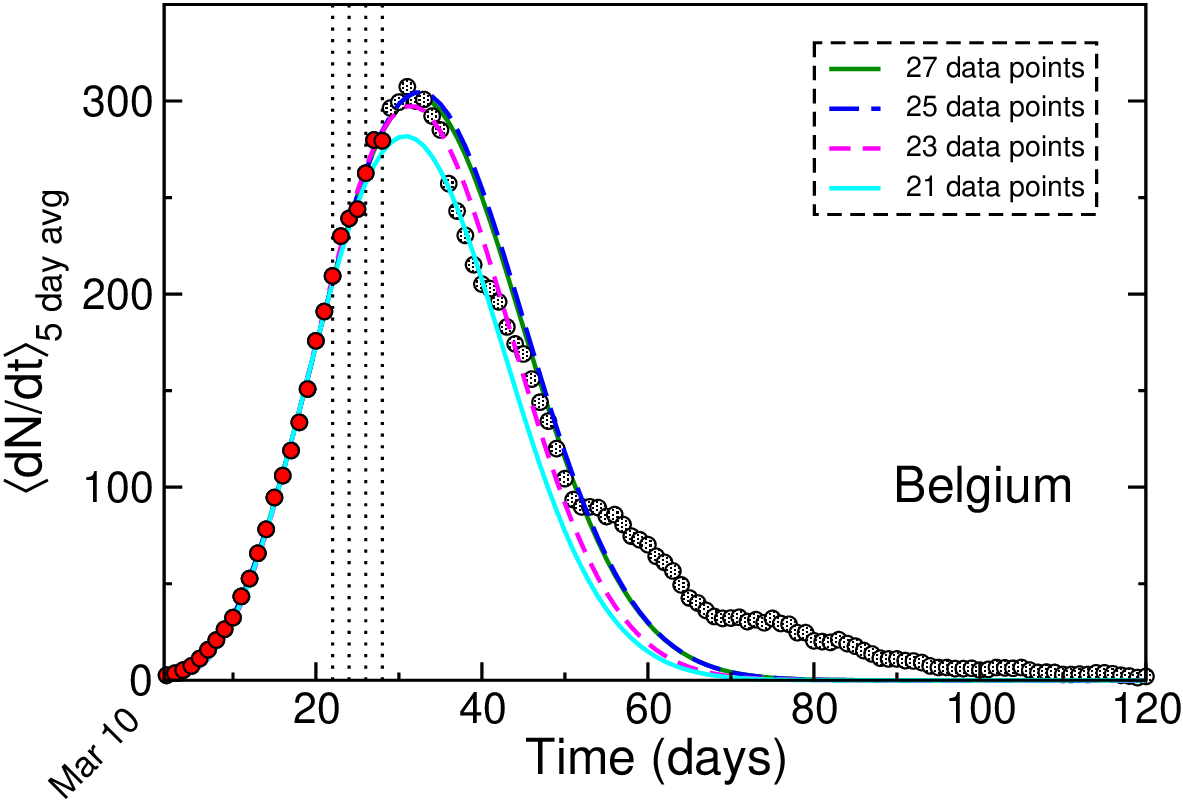}
}
\subfloat{
 \includegraphics[width=0.6\textwidth]{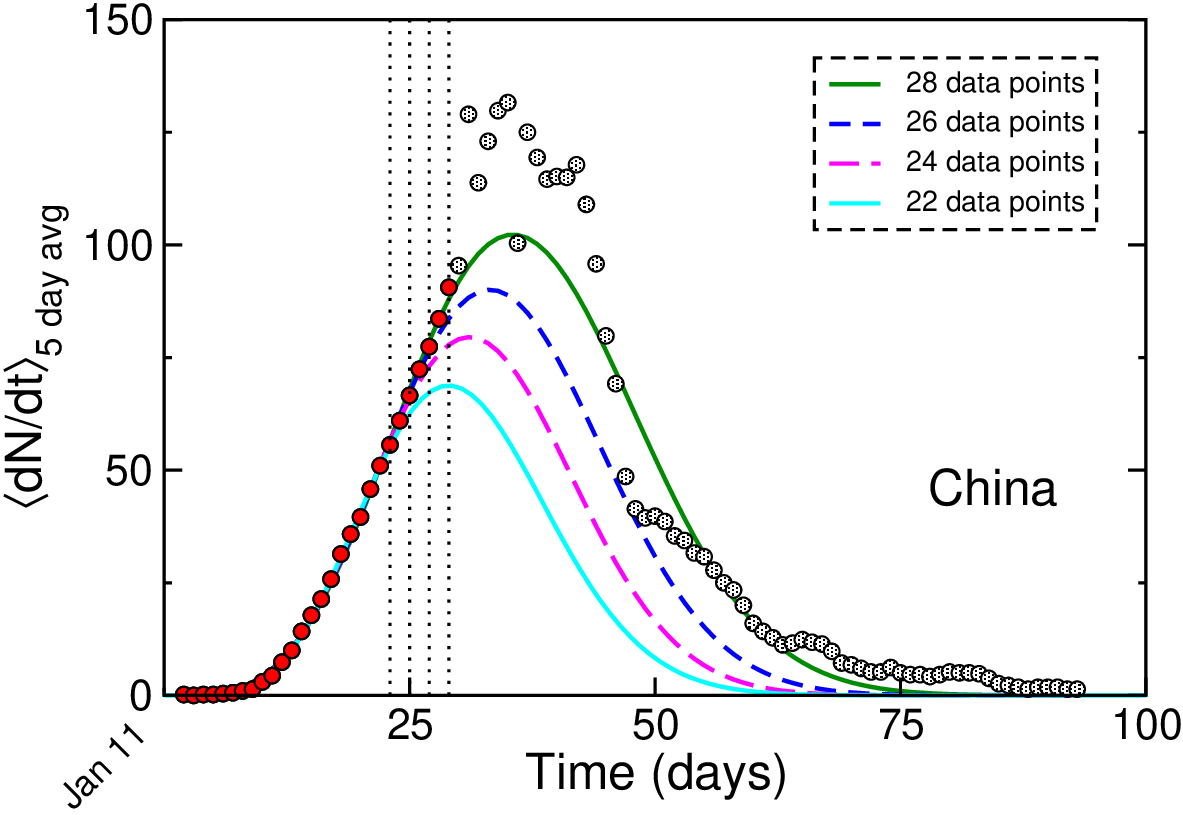}}
\vspace{0.5em}
\subfloat{
  \includegraphics[width=0.6\textwidth]{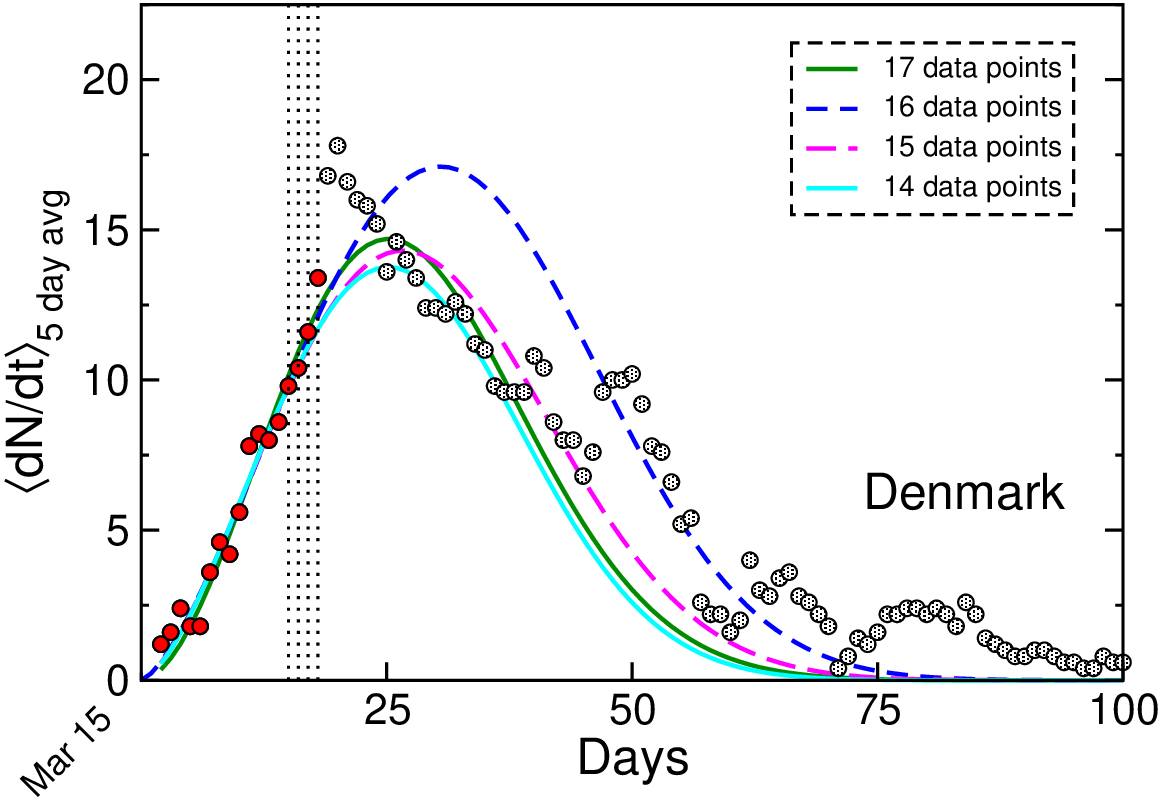}
  }
  \subfloat{
   \includegraphics[width=0.6\textwidth]{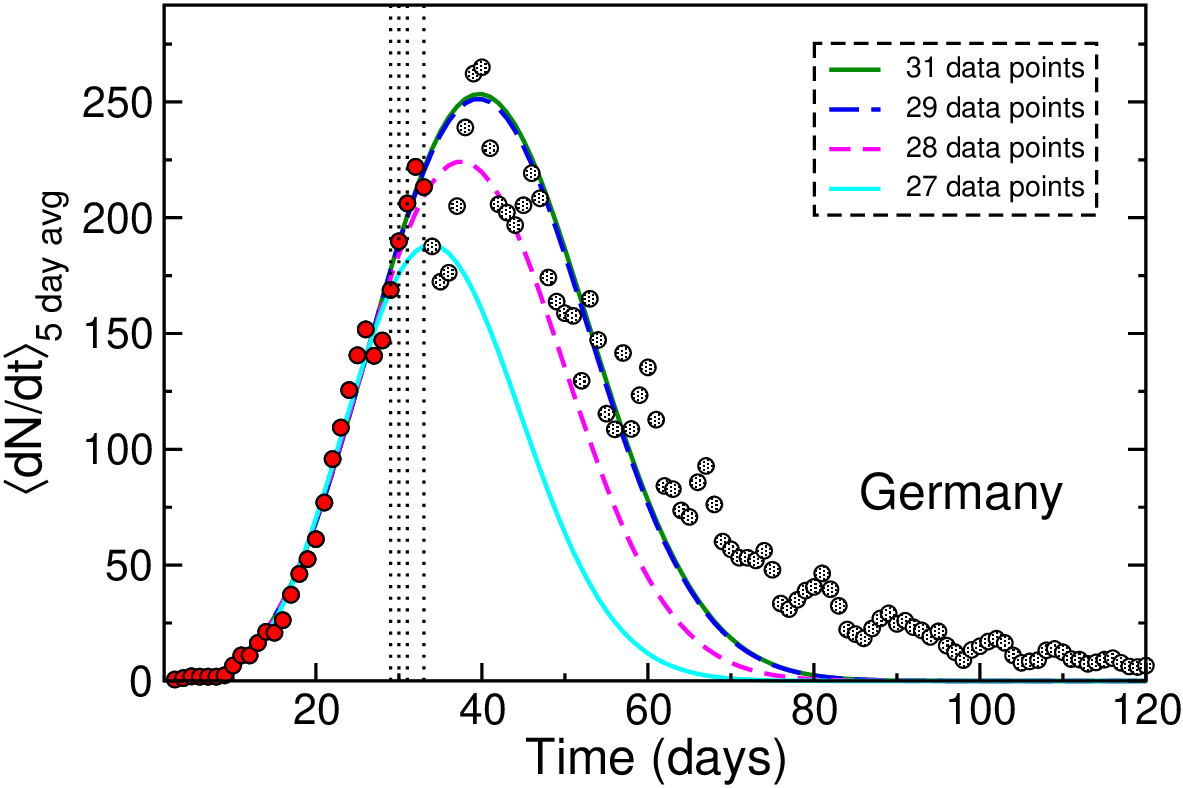}
   }
 \caption{\label{fig:calib} Power-law logistic growth models predictions using different ranges of available data in the rising part of the $dN/dt$ curve. The dotted lines mark the final in-sample calibration points from the data in Fig.~\ref{fig:rising}.  Similar to the other plots, the date of the first reported death (day 1) is indicated at $t = 0$ on the time axis.}
 \end{figure}
 \begin{figure}[htpb]
\subfloat{
  \includegraphics[width=0.51\textwidth]{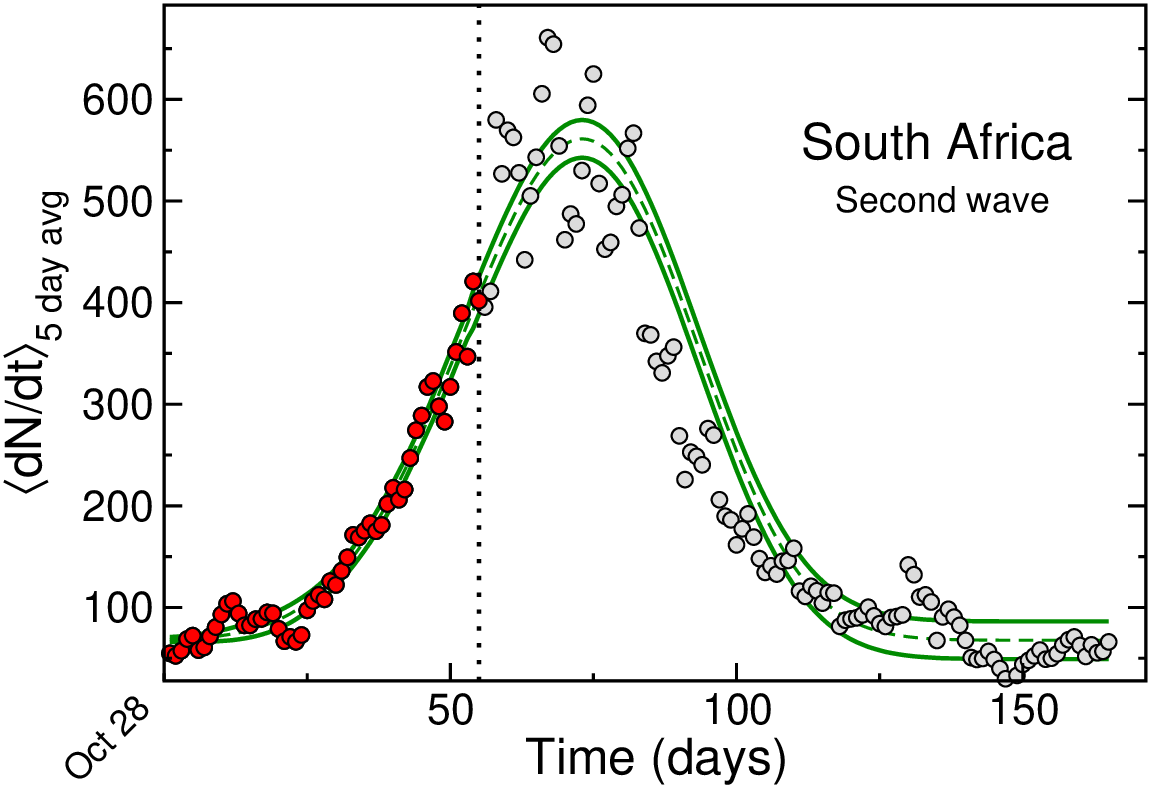}
  }
\subfloat{
 \includegraphics[width=0.51\textwidth]{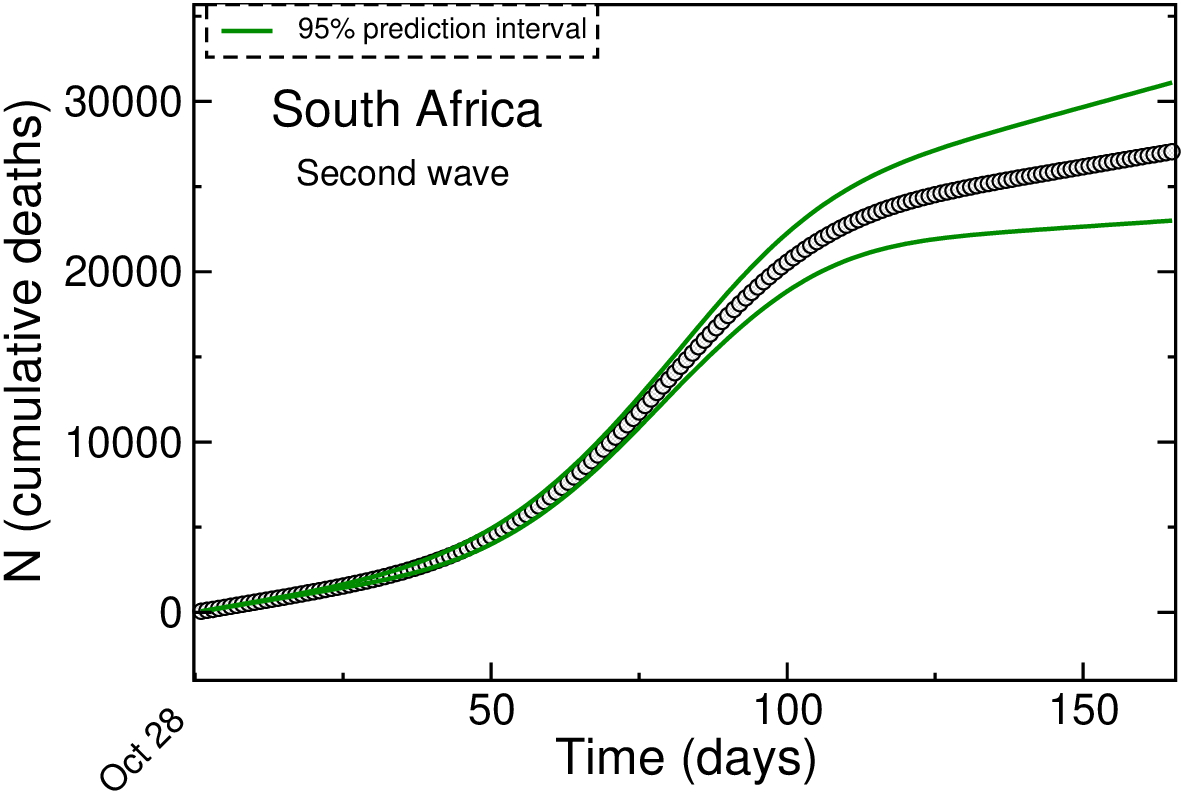}
 }
 \vspace{0.5em}
\subfloat{
  \includegraphics[width=0.51\textwidth]{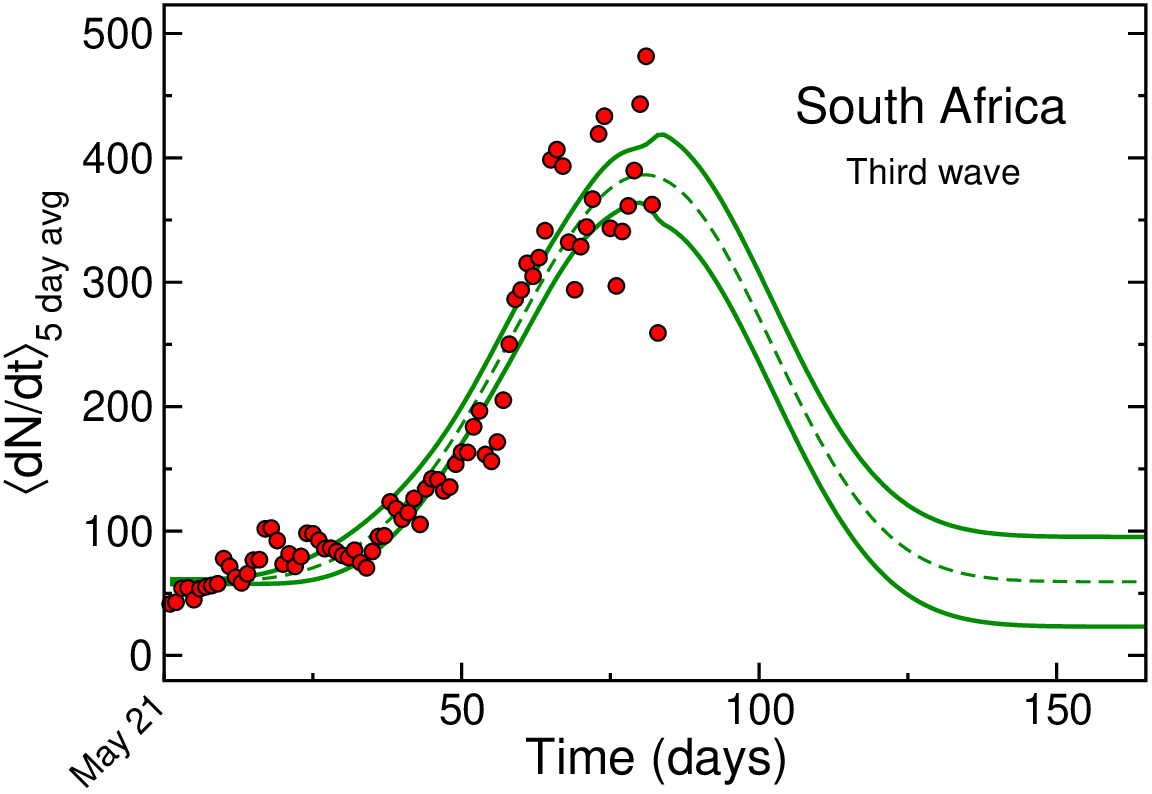}
  }
\subfloat{
 \includegraphics[width=0.51\textwidth]{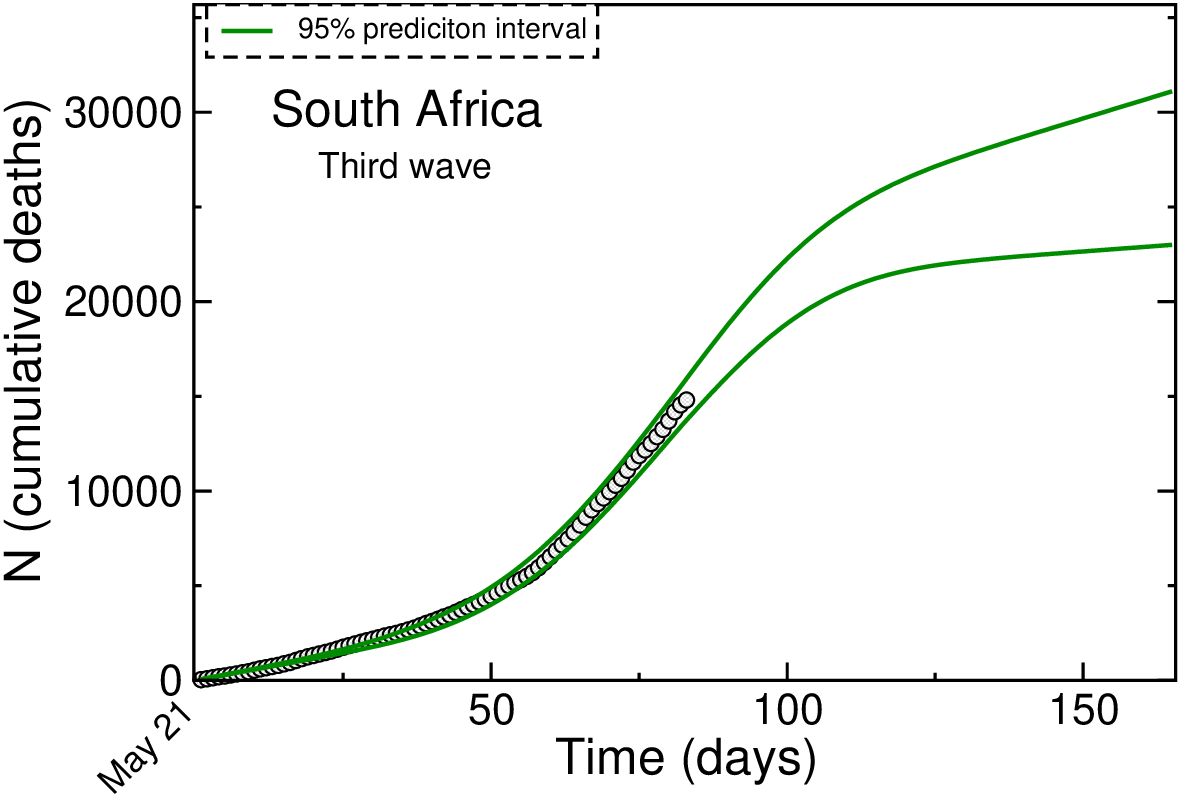}
 }
 
 \caption{\label{fig:final} Left panel: Power law LGM fits to second and third wave fatality data from South Africa. The data are obtained from~\cite{WHO2020} and are represented by filled circles. The start dates on the x-axis are for the years 2020 and 2021 respectively, for the second and third waves.  
 The red points show the in-sample calibration points used for the forecasting. The green curves show $\pm 95\%$ prediction intervals.  In each case, the date that marks the beginning of the data-analysis region for the epidemic wave is indicated at $t = 0$ on the time axis.   
 Right panel: Cumulative deaths recorded from the World Health Organization, shown together with our model forecasts at the 95\% prediction interval. }
 \end{figure}

 While this part of the analysis was necessary to show the more than satisfactory agreement of our power-law growth model with data, particularly from countries where the pandemic peak had long passed, this is not a robust test of the forecasting ability of the model. We next performed an out-of-sample forecasting test, this time \textit{only} using data points from the \textit{rising} part of the $dN/dt$ curve. To avoid fit convergences to `fake' minima that do not correspond to realistic values, we used a comparison with available data to empirically develop the procedure described below.  First the full parameter space was mapped to arrive in the vicinity of the correct SSR minimum. This was done by performing a first round of non-linear SSR minimization with the constraints\footnote{In some computer programs care should be taken that the fitted values for the parameters do not converge to their upper or lower bounds.} $0 < \lambda < 0.1$, $2 < \alpha < 3$, $2 < \gamma < 3$ and $0.5 < \epsilon < 1 $. After this initial step, the restrictions on $\lambda $, $\alpha$ and $\gamma$ were removed and the iterative grid-search NLS algorithm  was used evaluate the values that yielded the minimum SSR in that region of parameter space.  

 Our analyses showed that the procedure described above yielded reasonable agreement with the full data from different countries consistently. This agreement is illustrated in Fig.~\ref{fig:rising}, that also compares our fit results to those obtained from the classical LGM and both versions of the Richards LGM. The forecasting performance of the four models are further compared in Table~\ref{tab:table1}, which lists three  performance metrics. These include the root mean squared (RMS) error, 
\begin{equation}
{\rm RMSE} = \sqrt{\left(\frac{{\rm SSR}}{n}\right)},  
\end{equation}
where $n$ is the total number of data points in the curve, the mean absolute percentage error (MAPE),
\begin{equation}
{\rm MAPE} = \frac{1}{n}\sum_{i = 1}^{t_\textrm{days}} \left(\frac{|y(t_i)-D_i|}{D_i}\right)\times 100 
\end{equation}
and the coverage of the 95\% prediction interval, which quantifies the proportion of observations that fell within that range. 
As evident in the Fig.~\ref{fig:rising} and Table~\ref{tab:table1}, the predictions of the classical LGM and the generalized Richards model yield least agreement with the full data, while the Richards model agrees with some of the data.  The power-law growth model is found to be \textit{consistent} in its forecasting performance, showing reasonable agreement with the data in all cases. Fig.~\ref{fig:rising} also shows cumulative fatality data for each of the four countries, together with predictions of the power-law model. The power-law function also yielded consistently better agreement for the cumulative data. 
 
Another check of the robustness of our analysis was performed by systematically reducing the number of in-sample calibration points (marked in red in Fig.~\ref{fig:rising}) in the NLS fitting procedure. This effectively tested the stability of our predictions. The results of this systematic check are shown in Fig.~\ref{fig:calib} and Table~\ref{tab:table2}. The latter lists the number of in-sample data points used for the fits in each case, and the corresponding RMSE values as a forecasting metric. It is evident that as long as a reasonable number of `in-sample' data points are used, the power-law model makes reliable predictions. 

\subsection{Results for second and third wave data from South Africa (2020/2021)}
Once assured that our data analysis procedure was on a secure footing, we performed a similar analysis for the second and third waves in South Africa. It may be noted that strict lock down and containment policies were not imposed in these scenarios (compared to the first wave) and that only a partial $dN/dt$ curve for the third wave is available at the present time, allowing us to make predictions. Furthermore, the vaccinated status of part of the population and the different variants of the SARS-CoV-2 virus add additional complications that allow a rigorous test of the power-law growth model.   

Figure~\ref{fig:final} shows power-law growth model fits to the second and third wave fatality data from South Africa. The in-sample red data points were fit similarly as before, with two minor differences. Since the growth exponent is expected to be higher in these scenarios~\cite{st}, the ranges on $\alpha$ and $\gamma$ were increased to 2--6 in the initial restricted fit. An additional `background' parameter was required to added to Eq.~\eqref{eq:ours2}. This parameter took into account the roughly constant number of deaths/infections that occur between the waves. For the second wave, we observe excellent agreement between the data and the model predictions, which further validates the power-law LGM. The model also provides good agreement with cumulative fatality data obtained from the reported number of deaths during this time.  
Based on this validation, we use the power-law fit to make forecasts on the partial third wave epidemic curve for South Africa. The fit indicates that we are presently near the peak of the third wave for Covid-19 induced fatalities in South Africa. Similarly, the corresponding cumulative data indicate a starting of the flattening of the growth curve. 
 
\begin{sidewaystable}[htpb]
\centering
\caption{Out-of-sample forecasting performance metrics for the four countries, using different LGM models described in this work.}
\label{tab:table1}
\begin{tabular}{llccc}
\hline
\hline
\multicolumn{1}{c}{{Country}}&{\rm Growth }&\multicolumn{1}{c}{\rm Root mean}&\multicolumn{1}{c}{\rm Mean absolute } 
&\multicolumn{1}{c}{\rm{Percentage coverage of}}\\
 & model &\multicolumn{1}{c}{squared error}& \multicolumn{1}{c}{percentage error} & \multicolumn{1}{c}{95\% prediction interval}  \\
\hline
& Logistic& 51.7 & 76.4& 17 \\
Belgium& Richards & 17.4 & 25.8&76 \\
&Gen.  Richards& 31.8 & 64.1 & 21\\
& Power-law & 18.8 & 56.9& 60\\
\hline
& Logistic&28.8  &112.7 & 40\\
China& Richards &22.7  &175.2 & 32\\
&Gen.  Richards& 109.3 & 1679.5 & 31\\
& Power-law &11.4  & 40.8& 58\\
\hline
& Logistic& 3.9 & 78.0 & 61\\
Denmark& Richards &  2.6& 53.6 & 73\\
&Gen.  Richards& 24.2 & 3434.0& 15\\
& Power-law &2.0  &64.7 & 88\\
\hline
& Logistic& 57.1 & 85.7 & 26\\
Germany& Richards&  22.6 &28.0 & 59\\
&Gen.  Richards& 90.1 &465.1 & 25\\
& Power-law & 25.1 &56.8 & 47\\
\hline
\hline
\end{tabular}
 \end{sidewaystable}
\begin{table}[htpb]
\centering
\caption{Stability tests for power-law LGM forecasts, performed by a systematic removal of in-sample data points.}
\label{tab:table2}
\begin{tabular}{cccc}
 \hline
 \hline
\multicolumn{1}{c}{{Country}}&\multicolumn{1}{c}{{Number of}}&Color in&\multicolumn{1}{c}{{RMSE}}\\
&\multicolumn{1}{c}{{data points}}& Fig.~\ref{fig:calib}&\multicolumn{1}{c}{{(full data)}}\\
\hline
& 27 & Green & 18.4\\
Belgium& 25 & Blue& 19.1\\
& 23 & Magenta& 19.5\\
& 21 & Cyan & 22.2\\
\hline
& 28 & Green &10.0 \\
China& 26 & Blue& 16.6\\
& 24 & Magenta& 24.5\\
& 22 & Cyan & 31.7\\
\hline
& 17 & Green & 2.1\\
Denmark& 16 & Blue& 2.7\\
& 15 & Magenta& 1.8\\
& 14 & Cyan & 2.2\\
\hline
& 31 & Green & 24.8 \\
Germany& 29 & Blue& 24.7 \\
& 28 & Magenta& 30.5\\
& 27 & Cyan & 50.1 \\
\hline
\end{tabular}
 \end{table}
 \section{Summary}
 In summary, we used an empirical analysis to present a new logistic power-law growth model (LGM) that was applied to COVID-19 fatality data. This is relevant, as sub-exponential power-law growth is not adequately described by earlier variants of LGMs. Our model is found to be rather robust in accurately predicting peak and saturation values in epidemic growth curves from Belgium, China, Denmark and Germany. Following this validation, the power-law LGM is used to predict the COVID-19 induced-fatalities in the second and third waves for South Africa, after robustly testing the model predictions for the former.  We anticipate that our presented growth model will be useful for forecasting COVID-19 induced infections/deaths in other regions and epidemic spread in general.   

 \section{Acknowledgements}
 \label{ack}
 We are thankful to Prof. Niranjan Barik for useful discussions related to this work. ST acknowledges support from the NRF (National Research Foundation), South Africa, under SARChI Grant. No. 85100. 

\bibliography{pl_lgm_resubmit}
\end{document}